\documentclass[preprint]{aastex}
\usepackage{psfig}
\slugcomment{KSUPT-00/3 \hspace{0.5truecm} August 2000}

\newcommand{\omegal}{\Omega_{\Lambda}}
\newcommand{\kmsmpc}{{\rm \, km\, s}^{-1}{\rm Mpc}^{-1}}

\begin{document}

\title{Cosmological-Model-Parameter Determination from
Satellite-Acquired \\
Supernova Apparent Magnitude versus Redshift Data}

\author{Silviu Podariu\altaffilmark{1},
        Peter Nugent\altaffilmark{2},
        and
        Bharat Ratra\altaffilmark{1}
       }

\altaffiltext{1}{Department of Physics, Kansas State University,
Manhattan, KS 66506.}
\altaffiltext{2}{Lawrence Berkeley National Laboratory, MS50-232, 1
Cyclotron Road, Berkeley, CA 94720.}

\begin{abstract}
We examine the constraints that satellite-acquired
supernova apparent magnitude versus redshift data will place on
cosmological model parameters in models with and without a constant or
time-variable cosmological constant $\Lambda$. Data which
could be acquired in the near future will result in tight constraints
on these parameters. For example, if all other parameters of a
spatially-flat model with a constant $\Lambda$ are known, the
supernova data should constrain the non-relativistic matter density
parameter $\Omega_0$ to better than 1\% (2\%, 0.5\%) at 1$\sigma$ with
neutral (worst case, best case) assumptions about data quality.
\end{abstract}


\section{Introduction} 

Recent applications of the apparent magnitude versus redshift test
based on Type Ia supernovae (SNe~Ia) have resulted in interesting
constraints on cosmological-model parameters (see, e.g., Riess et
al. 1998; Perlmutter et al. 1999; Podariu \& Ratra 2000; Waga
\& Frieman 2000; Gott et al. 2001).  Higher quality
data will result in tighter constraints on cosmological-model
parameters. A dedicated supernova space telescope could provide the
high quality data needed to realize the full potential of this
neoclassical cosmological test.

In this paper we examine constraints on cosmological-model parameters
that will result from such a data set. For definiteness we focus on
data that could be acquired by the proposed SNAP space telescope
(Curtis et al. 2000 and http://snap.lbl.gov). That is, we assume a
data set of 2000 SNe~Ia multi-frequency light curves, for SNe
out to redshift $z = 2$, with errors discussed below.

Observational data favor models with a low $\Omega_0$. The simplest
such models have either flat spatial hypersurfaces and a constant or
time-variable cosmological ``constant" $\Lambda$ (see, e.g., Peebles
1984; Peebles \& Ratra 1988; Sahni \& Starobinsky 2000; Steinhardt
1999; Carroll 2000; Bin\'etruy 2000), or open spatial hypersurfaces
and no $\Lambda$ (see, e.g., Gott 1982, 1997; Ratra \& Peebles 1994,
1995; Kamionkowski et al. 1994; G\'orski et al.  1998). For a constant
$\Lambda$ (with density parameter $\omegal$), these models lie along
the lines $\Omega_0 + \omegal = 1$ and $\omegal = 0$, respectively, in
the more general two-dimensional ($\Omega_0$, $\omegal$)
model-parameter space. Depending on the values of $\Omega_0$ and
$\omegal$, models in this two-dimensional parameter space have either
closed, flat, or open spatial hypersurfaces. In this paper we derive
constraints on the parameters of the two-dimensional model as well as
those of the special one-dimensional cases.

We also derive constraints on the parameters of a spatially-flat model
with a time-variable $\Lambda$. The only known consistent model for a
time-variable $\Lambda$ is that based on a scalar field ($\phi$) with
a scalar field potential $V(\phi)$ (Ratra \& Peebles 1988). In this
paper we focus on the favored model which at low $z$ has $V(\phi)
\propto \phi^{-\alpha}$, $\alpha > 0$ (Peebles \& Ratra 1988; Ratra \&
Peebles 1988)\footnote{ Such a scalar field potential is present in
some high energy particle physics models (see, e.g., Rosati 2000;
Copeland, Nunes, \& Rosati 2000; Brax \& Martin 2000). Fujii (2000), 
Cormier \& Holman (2000), Faraoni (2000), Baccigalupi, Perrotta, \& 
Matarrese (2000), Dodelson, Kaplinghat, \& Stewart (2000), Ziaeepour 
(2000), Kruger \& Norbury (2000), Joyce \& Prokopec (2000), Goldberg 
(2000), Hebecker \& Wetterich (2000), Ure\~na-L\'opez \& Matos (2000), 
and Armendariz-Picon, Mukhanov, \& Steinhardt (2000) discuss this model
and other options.}.  This model is in reasonable accord with
observational data (see, e.g., Peebles \& Ratra 1988; Ratra \& Quillen
1992; Podariu \& Ratra 2000; Waga \& Frieman 2000; Brax, Martin, \&
Riazuelo 2000)\footnote{ See, e.g., Vishwakarma (2000), Ng \&
Wiltshire (2001), and Lima \& Alcaniz (2000) for
observational constraints on related models.}.

A scalar field is mathematically equivalent to a fluid with a
time-dependent speed of sound (Ratra 1991), and it may be shown that
with $V(\phi) \propto \phi^{-\alpha}$, $\alpha > 0$, the $\phi$ energy
density behaves like a cosmological constant that decreases with
time. We emphasize that in our analysis of this model here we do not
make use of the time-independent equation of state fluid approximation
to the model that has sometimes been used for such computations (see
the discussion in Podariu \& Ratra 2000).

Huterer \& Turner (1999), Starobinsky (1998), Nakamura \& Chiba
(1999), Saini et al. (2000), and Chiba \& Nakamura (2000) discuss
using supernova apparent magnitude versus redshift data to determine
the scalar field potential of the time-variable $\Lambda$ model. This
is a difficult task. Maor, Brustein, \& Steinhardt (2001) note that
even data of the quality anticipated from SNAP will not result in very
tight constraints on an arbitrary equation of state. They consider a
simple illustrative example, with an equation of state parameter $w$
that has two terms, one constant and the other linear in $z$. Maor et
al. show confidence contours (in a two-dimensional plane) for the two
parameters in the equation of state for this model in their Figure
2. After marginalizing over $\Omega_0$ the peak-to-peak spread in
their 2 $\sigma$ contour for the equation of state at $z = 0$, $w_0$,
is about $-0.3$ for $w_0 = -0.7$, or about 43\% of the value of
$w_0$. This corresponds to a symmetrized 2 $\sigma$ uncertainty of
about $\pm 22\%$ on $w_0\,$\footnote{ We acknowledge helpful
discussions with P. Steinhardt on this issue.}.  The corresponding
peak-to-peak spread in their 1 $\sigma$ contour is about $-0.22$,
which corresponds to a 1 $\sigma$ uncertainty of about $\pm 16\%$ on
$w_0$. For fixed $\Omega_0$, the peak-to-peak spread in their 1
$\sigma$ contour is about $-0.09$, which corresponds to a 1 $\sigma$
uncertainty of about $\pm 6.5\%$ on $w_0$. While much larger than the
constraints we place on model-parameter values (see below) this is
still a reasonably precise determination of $w_0$.

Motivated by the approach adopted in analyses of current supernova
apparent magnitude versus redshift data (see, e.g., Riess et al. 1998;
Perlmutter et al. 1999), we instead focus on how well future supernova
data will constrain parameters of various cosmological
models\footnote{ A similar approach is used in analyses of cosmic
microwave background anisotropy data. Here one computes predictions of
a theoretical model as a function of a few cosmological parameters and
derives constraints on these parameters by comparing these predictions
to observational data, either using an approximate $\chi^2$ technique
(see, e.g., Ganga, Ratra, \& Sugiyama 1996; Dodelson 2000; Le Dour et
al. 2000; Lange et al.  2001; Balbi et al. 2000), or using the
complete models-based maximum likelihood technique (see, e.g., Ganga
et al. 1997, 1998; Ratra et al.  1998, 1999; Rocha et al. 1999).}.

We want to determine how well supernova data distinguishes between
different cosmological-model-parameter values. To do this we pick a
model and a range of model-parameter values and compute the luminosity
distance $D_{\rm L} (z)$ for a grid of model-parameter values that
span this range.  Figure 1 shows examples of $D_{\rm L} (z)$'s
computed in the time-variable $\Lambda$ model (Peebles \& Ratra 1988).

The error bars on the supernova fluxes are the ones that are most
likely to be symmetric (and thus allow for the simplest comparison
between model predictions and observational data), so we work with
flux $f \propto D_{\rm L}{}^{-2}$ for the comparison between model
predictions and anticipated data. For our purposes, the constant of
proportionality in this relation is unimportant since the SNe in the
final reduced data set have been made standardized candles (see, e.g.,
Phillips 1993, and more recently Riess et al. 1998; Perlmutter et
al. 1999).

For computational simplicity we assume supernova data from SNAP will
be combined to provide fluxes and errors on fluxes for 67 uniform bins
in redshift, of width $\Delta z = 0.03$, with the first one centered
at $z = 0.03$ and the last one at $z = 2.01$. In each bin the
statistical and systematic errors are combined to give a flux error
distribution with standard deviation $\sigma (z)$. 

To determine how well supernova data will distinguish between
different sets of model-parameter values, we pick a fiducial set of
model-parameter values which give a flux $f_{\rm F}(z)$ and compute
\begin{equation}
  N_\sigma (P) = \sqrt{ \sum_{i = 1}^{67} \left({f(P, z_i) - f_{\rm F}
      (z_i) \over \sigma(z_i) f_{\rm F} (z_i) } \right)^2 } ,
\end{equation}
where the sum runs over the 67 redshift bins and $P$ represents the
model parameters, for instance $\Omega_0$ and $\omegal$ in the general
two-dimensional constant $\Lambda$ case. $N_\sigma (P)$ is the number
of standard deviations the model-parameter set $P$ lies away from that
of the fiducial model. This representation (eq. [1]) is exact 
for the case where the correlated errors between redshift bins for the
distance determinations are negligible. 

The error budget is summarized in the next section. Results are 
presented and discussed in $\S$3 and we conclude in $\S$4.

\section{Error Budget}

The following provides a brief overview of the constraints that a
satellite-based supernova program can place on both the statistical
and many of the potential systematic errors. For a more complete
discussion see the SNAP proposal.

\subsection{Statistical Errors}

Currently a single SN~Ia provides a $\approx$ 16\% measurement of the
flux ($\approx$ 8\% in distance) (Jha et al. 1999). A large fraction
of this uncertainty almost certainly resides in the correction for
extinction. By going to space one will be able to greatly increase the
wavelength coverage and precision of the photometric measurements,
thereby reducing this uncertainty considerably. (Signal-to-noise of $\sim 30$
could be achieved for a SN at AB(1.0$\mu$m) = 27.0, with systematics in the
absolute photometry of $<$1\%.) The SNAP satellite has
baselined 15 broad-band filters from about 0.3 to 1.7 $\mu$m in
addition to obtaining spectrophotometry near peak for each SN~Ia. A
conservative estimate of the intrinsic uncertainty for a given SN~Ia
with this type of data set would be $\approx$ 10\% in flux ($\approx$
5\% in distance). There is potential for reducing this even further
through the identification of additional parameters that constrain the
corrected peak luminosity of SNe~Ia beyond the single parameter of
light-curve shape currently used. Here we will conservatively assume
that the statistical uncertainty in satellite-based SN~Ia measurements
such as these will be 10\% in flux. $\sqrt{N}$ statistics on 2000 SNe~Ia
over the 67 aforementioned bins would provide an uncertainty of $<2$\%
per bin.

\subsection{Systematic Errors}

A major advantage of a space telescope is the much better opportunity
for controlling (or studying) the many known (and unknown) sources of
error. These include environmental effects, evolution, intergalactic
dust, unusual cases which bias the distribution, etc. See, e.g.,
Howell, Wang, \& Wheeler (2000), Aldering, Knop, \& Nugent (2000),
Croft et al. (2000), Nomoto et al. (2000), Barber (2000), Hamuy et
al. (2000), Livio (2000), Totani (2000), and Gott et al. (2001) for
discussions of some of these issues.  Without understanding and
limiting these sources of error an accurate measurement of the
cosmological parameters can not be obtained. Here we mention a few of
the potential sources of systematic errors and how space-based observations
could constrain or eliminate them (a more
detailed discussion of these and other sources of systematic errors
can be found in the SNAP proposal).

{\it Malmquist Bias}. This is the sampling bias due to any 
low-versus-high-redshift difference in detection efficiency of
intrinsically fainter supernovae.  For the aforementioned redshift
range, the proposed experiment will attempt to detect every
supernova in the observed region of sky at 10\% of its peak
brightness, thus eliminating this source of systematic uncertainty.

{\it Extinction by ``Ordinary'' Dust}. The proposed experiment
will attempt to obtain cross-wavelength-calibrated data with broad wavelength
coverage for each supernova, so that the dimming of the spectrum as a
function of wavelength can be measured with high signal-to-noise.
Furthermore, SNe~Ia in early-type galaxies with little to no
extinction will be targeted to precisely determine the intrinsic
colors of a SN~Ia at a variety of light-curve shapes (see Riess et
al. 1996 for a study of this at low redshift). This would then allow
one to study the ratio of selective to total absorption from dust and
correct for any potential evolution of this ratio as a function of
redshift.

{\it Extinction by ``Gray'' Dust}. It has been suggested by
Aguirre (1999) that certain large (up to $\sim$0.1\AA), and possibly
needle-like, dust grains can be expelled from galaxies via radiation
pressure and can have an opacity curve that is shallow in optical
bands, thus making them absorptive while producing only small color
excess. Such dust would lead the unwary cosmologist
into underestimating $\Omega_M$ or overestimating $\Omega_\Lambda$,
thus producing a systematic bias. If there is gray dust that has had
insufficient time to diffuse uniformly in intergalactic space,
different lines of sight would have differing amounts of extinction
due to clumping.  This would result in an increase of observed
supernova magnitude dispersion, an effect that is not seen in current
observations, and could easily be detected by a space-based experiment. 
Furthermore, it is also possible to detect $z < 0.5$ gray dust
by comparing optical and near-IR photometry of SNe (both Ia and II)
found in this redshift range since the dust is not completely gray and
will show a color excess over a large enough wavelength range
(see, e.g., Riess et al. 2000).

{\it SN~Ia Evolution}. SNe~Ia with different progenitor properties
should result in explosions with slightly differing properties, even
if there is only one mechanism for creating them, and even if this
mechanism has a set ``trigger'' such as the Chandrasekhar limit
(H\"oflich et al. 2000). If these differences are not corrected by
the light curve width-luminosity relation presently in use, and
if the distribution of key parameters of the progenitor stars
changes with redshift, the SN~Ia explosions observed at high redshift
could differ in peak luminosity from those at low redshift, leading to
a systematic error in the determination of the cosmological parameters. 
However, a dataset acquired from a space telescope should allow corrections 
for these differences, or allow similar SNe~Ia to be identified and
matched at high and low redshifts thus mitigating against the effects
of changing progenitor properties.

One of the wonderful aspects of using SNe~Ia for cosmology is the fact
that the supernova bares its entire history, from progenitor through
explosion, to the observer. Thus the supernova can't hide the effects
of evolution since these will make themselves apparent in the
light curves and spectra. Figure 2 illustrates this statement. This shows
the temporal spectral
evolution of a typical SN~Ia. At very early times one probes the
outer, unburned layers left over from the progenitor. As seen in
Fisher et al. (1997) this epoch displays spectral features from
high-velocity carbon left over from the original progenitor and could
be used to tightly constrain various theoretical models. At later
times, near peak brightness, we are beginning to probe the layers of
the atmosphere which show the intermediate-mass elements systhesized
in the runaway thermonuclear explosion. Nugent et al. (1995) showed how
some of the spectroscopic features of these elements (\ion{Si}{2} \&
\ion{Ca}{2}) nicely correlate with the peak brightness of the
SN~Ia. Finally, during the nebular phase, we note the strong
\ion{Fe}{2} emission lines at low velocity left over from the
radioactive decay of $^{56}$Ni to $^{56}$Co to $^{56}$Fe. These
observations allow one to directly probe the total amount of $^{56}$Ni
synthesized during the explosion (see Kuchner et al. 1994 and Fisher et
al. 1995).

Curtis et al. (2000) have identified a series of key observable
supernova features that reflect differences in the underlying physics
of the supernova. By measuring all of these features for each
supernova one should be able to tightly constrain the physical conditions of the
explosion, making it possible to recognize supernovae that have
similar initial conditions and/or arise in matching galactic
environments.  The current theoretical models of SN~Ia explosions are
not sufficiently complete to predict the precise luminosity of each
supernova, but they are able to give the rough correlations between
changes in the physical conditions of the supernovae and the peak
luminosity (H\"oflich, Wheeler, \& Thielemann 1998). These conditions
include the velocity of the ejecta (a measurement of the kinetic
energy of the explosion), the opacity of the
inner layers (which affects the overall light curve shape), the
metallicity of the progenitor (which affects the early spectra),
$^{56}$Ni mass (a measurement of the total luminosity), and $^{56}$Ni
distribution (which might lead to small effects in the light curve
shape at early time).  One can therefore give the approximate accuracy
needed for the measurement of each feature to ensure that the physical
condition of each set of supernovae is well enough determined so that
the range of luminosities for those supernovae is well below the
systematic uncertainty bound of $\sim$2\% when all the constraints are
used together (see Curtis et al. 2000 for a full description).

\section{Results and Discussion} 

For SNAP data $\sigma (z)$ (eq. [1]) is estimated to be
2\% in each redshift bin up to $z = 1.7$, and then increasing linearly
with redshift to 10\% at $z = 2$. This is the ``neutral" case. The
``best" case assumes that errors are limited by $\sqrt{N}$ statistics
(with systematic errors at or below the 1\% level), giving $\sigma (z)
= 1\%$ over the whole redshift range. The ``worst" case (this is the 
baseline SNAP mission) assumes $\sigma (z) = 3\%$ to
$z = 1.2$ and $=10\%$ from $z = 1.2$ to $z = 2$.

Figure 3 illustrates the ability of anticipated space telescope data to 
constrain
cosmological-model parameters for the general two-dimensional constant
$\Lambda$ case. SNAP data with even worst case error bars will lead to
greatly improved cosmological-parameter determination (see, e.g.,
Riess et al. 1998, Perlmutter et al. 1999, and Podariu \& Ratra 2000
for constraints from current data). We note that as expected the
contours are elliptical, indicating that one combination of the
parameters is better constrained than the other orthogonal combination
(see, e.g., Goobar \& Perlmutter 1995).

Figure 4 illustrates the ability of space telescope data to distinguish 
between a
constant and a time-variable $\Lambda$ in a spatially-flat model. The 
fiducial model here is a constant $\Lambda$ ($\alpha = 0$) model with 
$\Omega_0 = 0.28$ and $\omegal = 0.72$. SNAP data with even worst case 
error bars will result in greatly improved discrimination (see, e.g., 
Podariu \& Ratra 2000 for the current situation).  We note again that 
the contours are elliptical.

Figure 5 illustrates the ability of space telescope data to constrain $\Omega_0$
and $\alpha$ in the spatially-flat time-variable $\Lambda$ model
(Peebles \& Ratra 1988). Here the time-variable $\Lambda$ fiducial model 
has $\Omega_0 = 0.2$ and $\alpha = 4$. Again, SNAP will allow for tight 
constraints on these cosmological parameters.

If other data (such as cosmic microwave background anisotropy
measurements from MAP and Planck Surveyor and weak-lensing studies
from the proposed SNAP mission) pinned down some of the cosmological
parameters, the supernova data would then be able to provide tighter
constraints on the remaining parameters. For instance, Figure 6 shows
constraints from space telescope data on $\Omega_0$ in a spatially-flat constant
$\Lambda$ model and in an open $\Lambda = 0$ model. As expected from
the elliptical shape of the contours in Figure 3, anticipated
supernova data will constrain $\Omega_0$ more tightly in the
spatially-flat case than in the open case. In both cases SNAP will
provide tight constraints on $\Omega_0$. For instance, at 3 $\sigma$,
in the spatially-flat model we find $\Omega_0 = 0.3 \pm 0.007$, $= 0.3
\pm 0.015$, and $= 0.3 \pm 0.003$ for neutral, worst, and best case
errors, while in the open model we have $\Omega_0 = 0.3 \pm 0.015$, $=
0.3 \pm 0.03$, and $= 0.3
\pm 0.006$ for neutral, worst, and best case errors.

Figure 7 shows the space telescope data constraints on $\Omega_0$ and 
$\alpha$ in the spatially-flat time-variable $\Lambda$ model, if other data 
were to require 
that either $\alpha = 4$ or $\Omega_0 = 0.2$. SNAP data will provide tight
constraints on these parameters.  For instance, if $\alpha = 4$ we
find $\Omega_0 = 0.2 \pm 0.009$, $= 0.2 \pm 0.02$, and $= 0.2 \pm
0.004$ for neutral, worst, and best case errors, while if $\Omega_0 =
0.2$ we have $\alpha = 4 \pm 0.25$, $= 4 \pm 0.5$, and $= 4 \pm 0.1$
for neutral, worst, and best case errors, all at 3 $\sigma$.

\section{Conclusion}

Supernova space telescope data of the quality assumed here will lead
to tight constraints on cosmo\-logical-model parameters. For instance,
in a spatially-flat constant $\Lambda$ model where all other parameters are 
known, anticipated space telescope supernova data will determine $\Omega_0$
to about $\pm 0.8\%$, $\pm 1.7\%$, and $\pm 0.4\%$ (for neutral,
worst, and best case errors respectively) at 1 $\sigma$. The
corresponding errors on $\Omega_0$ for the open case are about $\pm
1.6\%$, $\pm 3.7\%$, and $\pm 0.7\%$. For the time-variable $\Lambda$
model, when $\alpha$ is fixed, $\Omega_0$ will be known to about $\pm
1.5\%$, $\pm 3.1\%$, and $\pm 0.7\%$, respectively, while when
$\Omega_0$ is fixed, $\alpha$ will be determined to about $\pm 2.2\%$,
$\pm 4.2\%$, and $\pm 1.2\%$.  This will have important consequences
for cosmology.

\bigskip

We acknowledge valuable discussions with G. Aldering, M. Levi, S. Perlmutter,
and T. Soura\-deep. SP and BR acknowledge support from NSF CAREER grant 
AST-9875031 and PN acknowledges computational support from the DOE Office of 
Science under Contract No. DE-AC03-76SF00098.

\begin{figure}[p]
\psfig{file=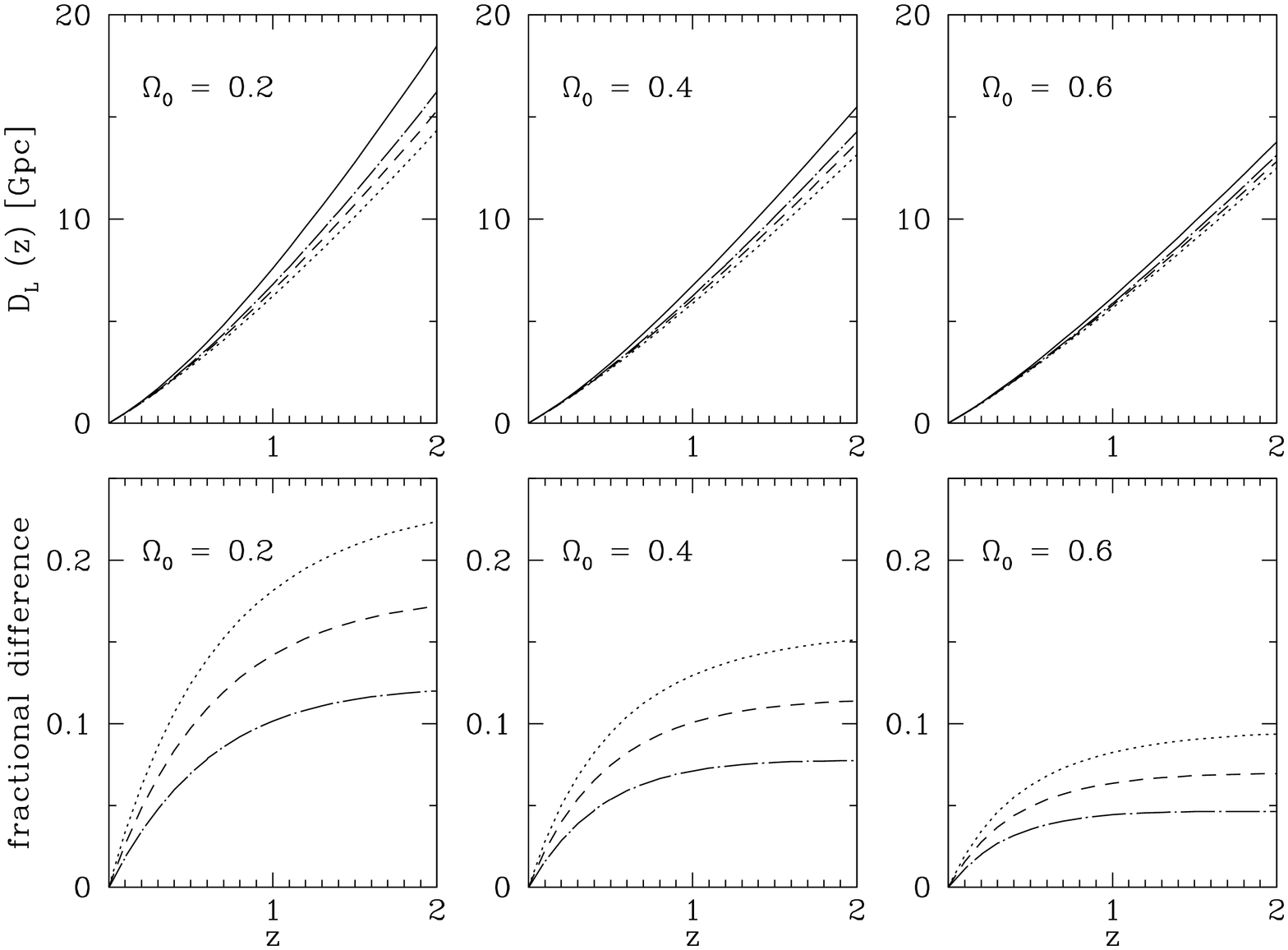,height=7.0in,width=7.0in,angle=270}
\caption{Lines in the panels in the upper row show luminosity
distance $D_{\rm L} (z, \alpha)$ as a function of redshift $z$ for
various values of $\alpha$ computed for Hubble parameter $H_0 = 65
\kmsmpc$ for the spatially-flat time-variable $\Lambda$ model with
scalar field potential $V(\phi) \propto \phi^{-\alpha}$. In descending
order at $z = 2$ the lines correspond to $\alpha$ = 0, 2, 4, and 8
(solid, dot-dashed, dashed, and dotted curves respectively). $\alpha =
0$ is the constant $\Lambda$ model. From left to right the three
panels correspond to $\Omega_0$ = 0.2, 0.4, and 0.6.  The three lower
panels show the fractional differences relative to the $\alpha = 0$
case, $1 - D_{\rm L}(z, \alpha)/D_{\rm L}(z, \alpha = 0)$, as a
function of $z$, for the values of $\Omega_0$ used in the upper
panels. Here the lines correspond to $\alpha$ = 8, 4, and 2, in
descending order at $z = 2$.}
\end{figure}

\begin{figure}[p]
\psfig{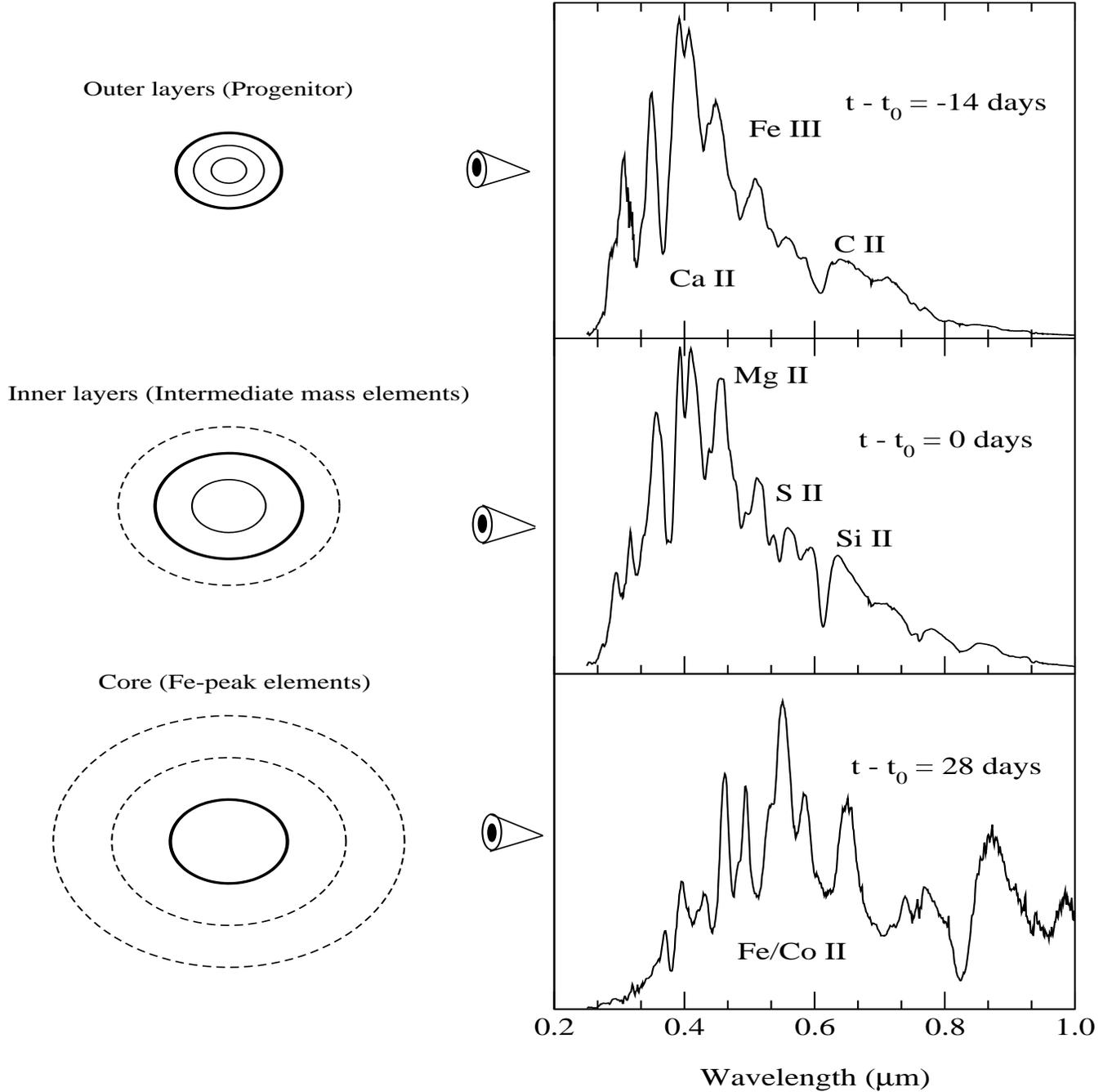}
\caption{As the SN~Ia evolves in time the rapid expansion of it's
atmosphere allows the observer to probe to deeper layers as the optical
depth falls off with the diminishing density. At early times one views the
outermost layers, mostly composed of the unburnt progenitor. Near peak
brightness the intermediate mass elements of \ion{S}{2} and \ion{Si}{2}
are quite visible. At later times, shortly after entering the nebular
phase, one views the Fe-peak core of the SN~Ia where the radioactive decay
of $\approx$0.5 M$_{\odot}$ of $^{56}$Ni has taken place since the
explosion.}
\end{figure}

\begin{figure}[p]
\psfig{file=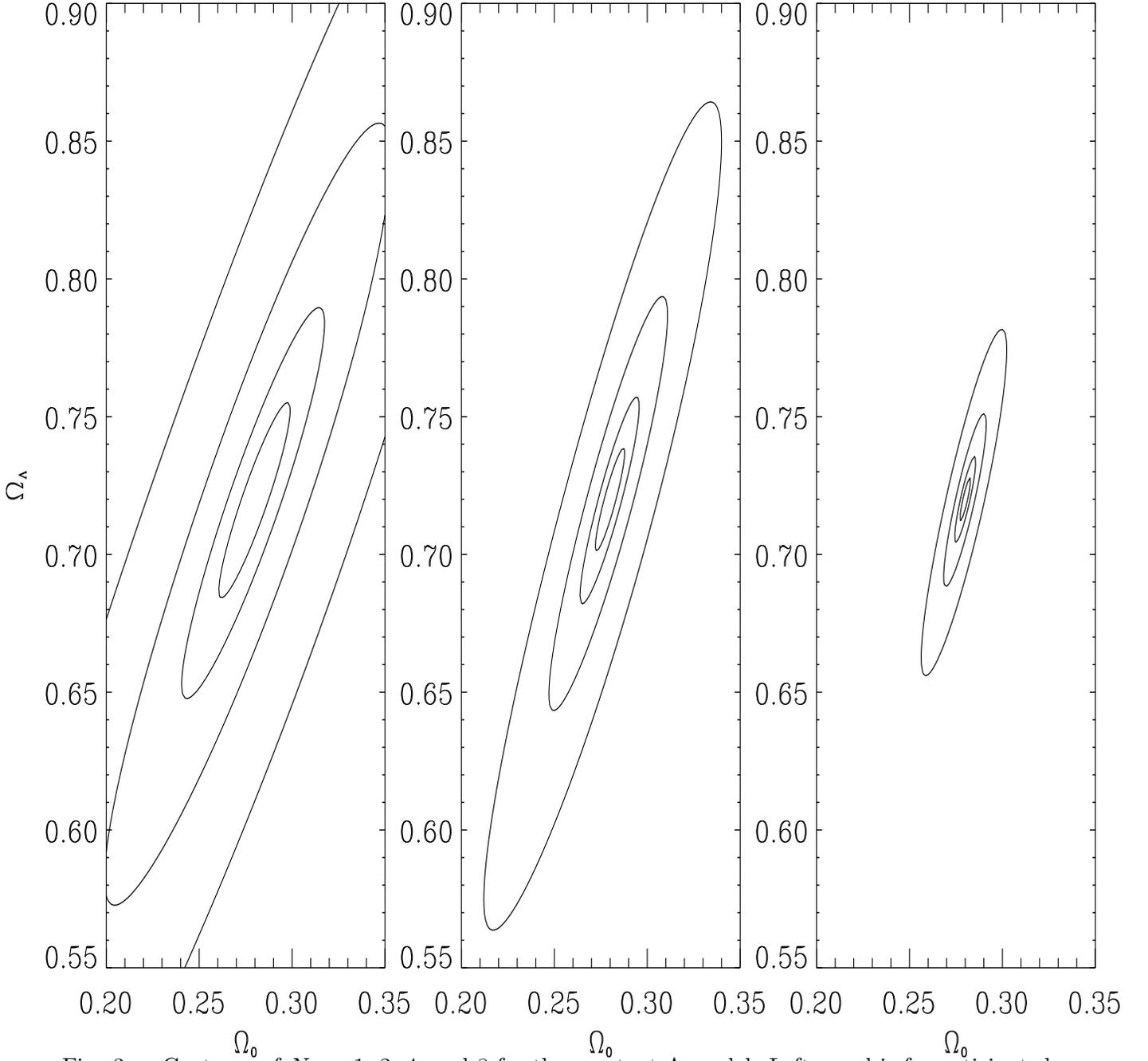,height=6.7in,width=7.0in,angle=90}
\caption{Contours of $N_\sigma$ = 1, 2, 4, and 8 for the constant
$\Lambda$ model. Left panel is for anticipated supernova data with worst
case errors, center panel is for neutral case errors, and right panel
is for best case errors. The fiducial model is spatially-flat with
$\Omega_0 = 0.28$ and $\omegal = 0.72$.}
\end{figure}

\begin{figure}[p]
\psfig{file=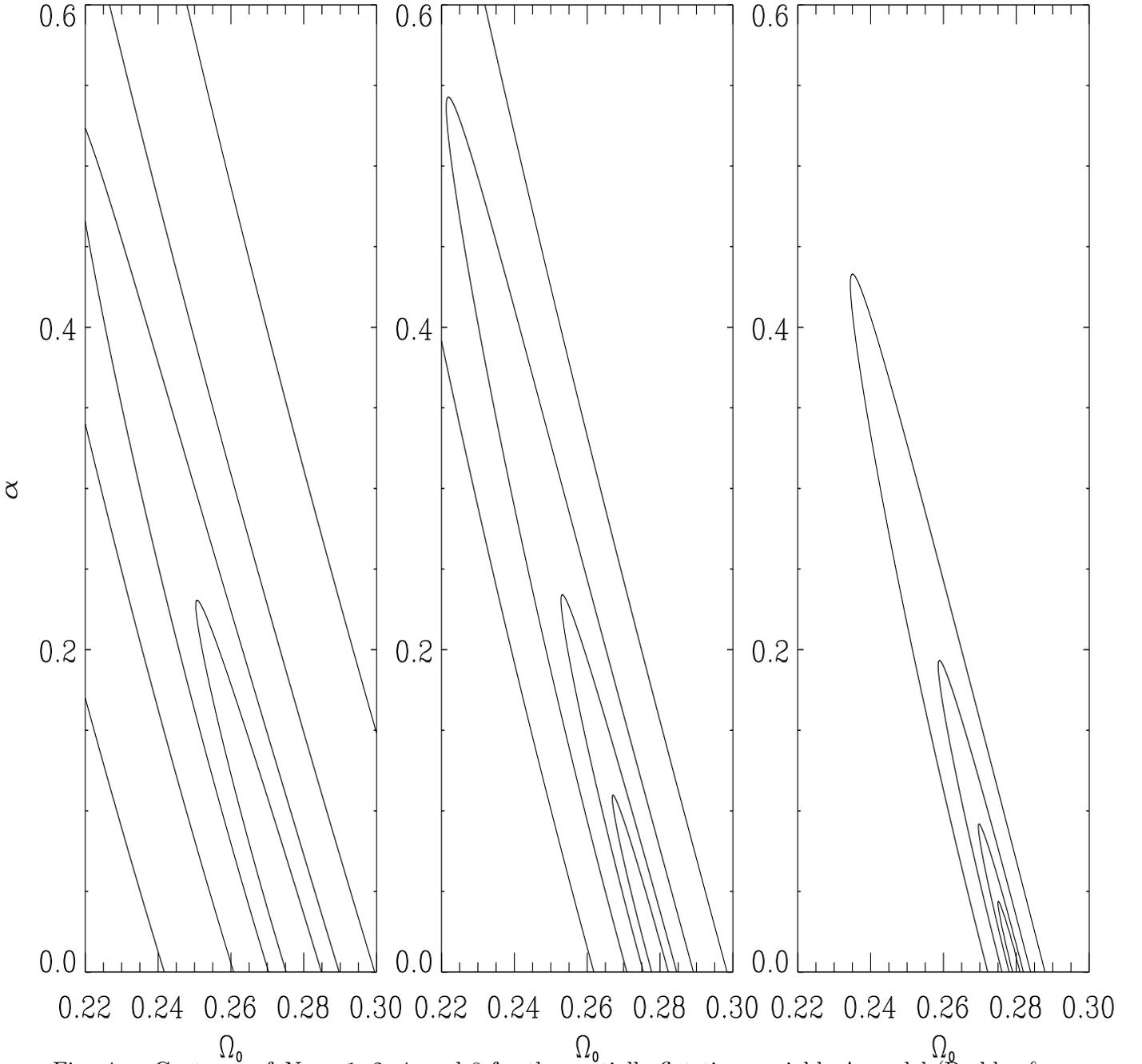,height=6.7in,width=7.0in,angle=90}
\caption{Contours of $N_\sigma$ = 1, 2, 4, and 8 for the
spatially-flat time-variable $\Lambda$ model (Peebles \& Ratra
1988). Left panel is for anticipated supernova data with worst case errors,
center panel is for neutral case errors, and right panel is for best
case errors. The fiducial model has $\Omega_0 = 0.28$ and $\alpha = 0$
(and is thus a constant $\Lambda$ model with $\omegal = 0.72$; this
was also the fiducial model used for Fig. 2).}
\end{figure}

\begin{figure}[p]
\psfig{file=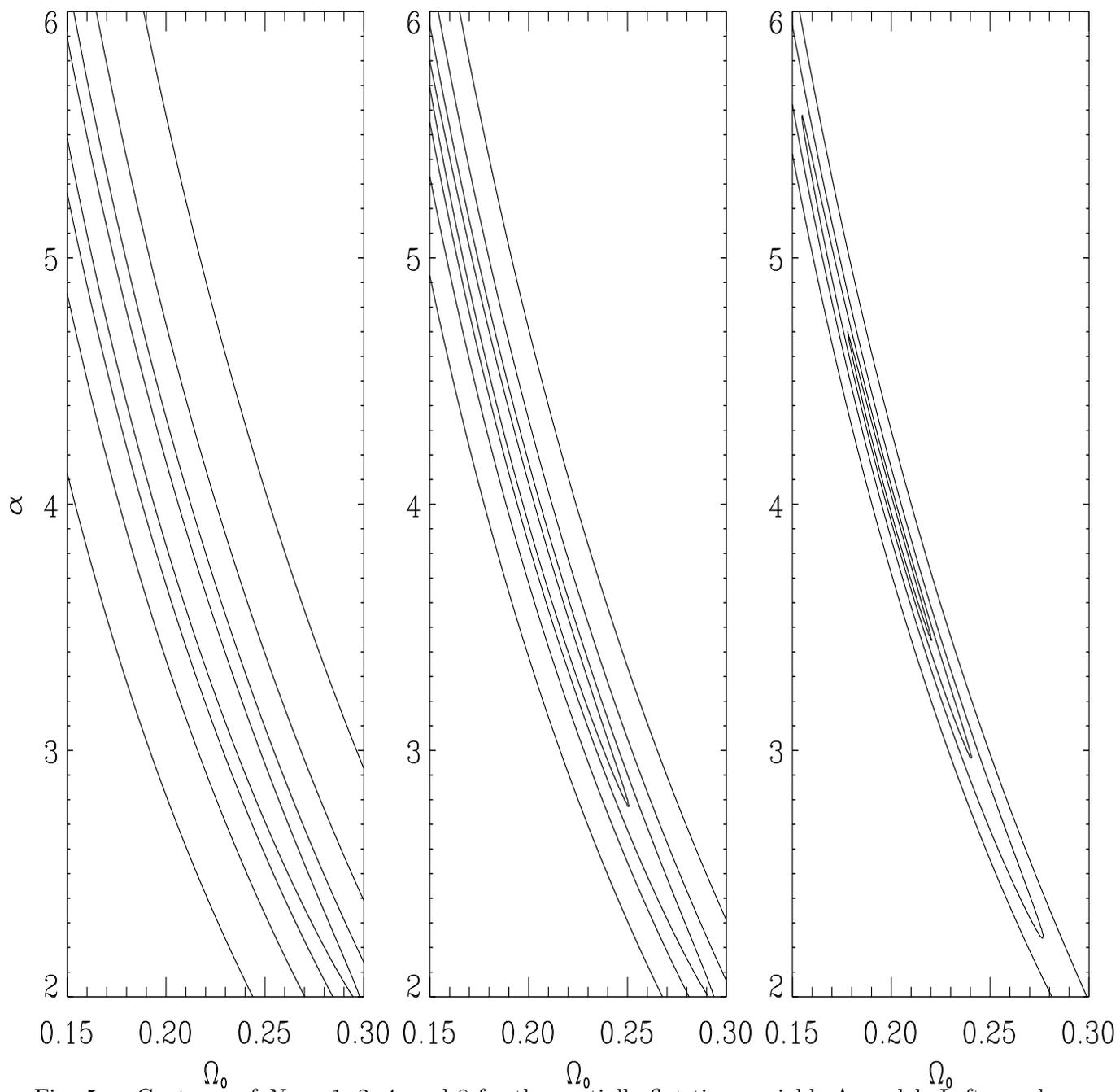,height=6.7in,width=7.0in,angle=90}
\caption{Contours of $N_\sigma$ = 1, 2, 4, and 8 for the
spatially-flat time-variable $\Lambda$ model. Left panel is for
anticipated supernova data with worst case errors, center panel is for
neutral case errors, and right panel is for best case errors. The
fiducial model has $\Omega_0 = 0.2$ and $\alpha = 4$.}
\end{figure}

\begin{figure}[p]
\psfig{file=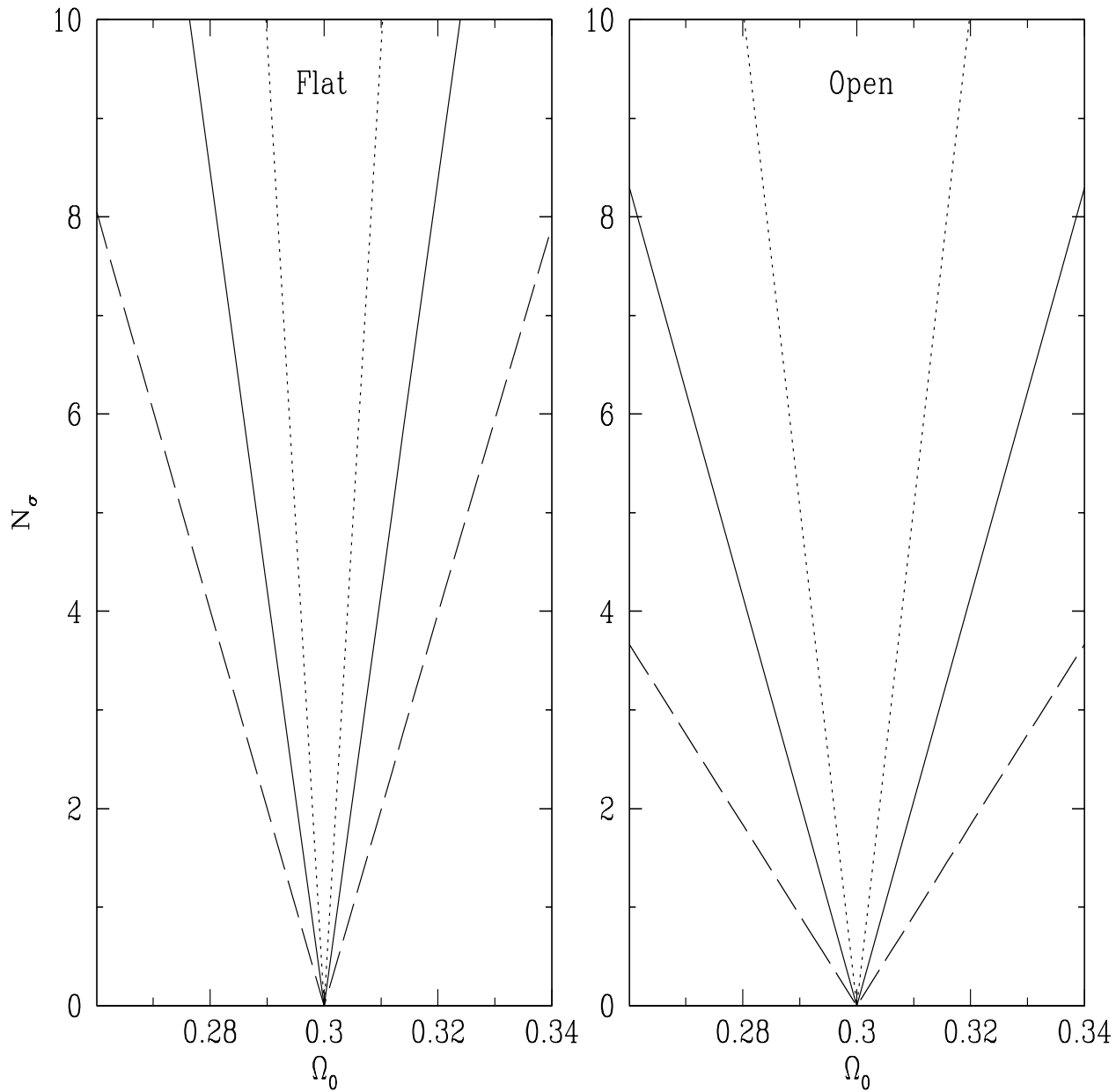,height=7.0in,width=7.0in,angle=270}
\caption{$N_\sigma (\Omega_0)$ for a flat model with a constant
$\Lambda$ (left panel) and for an open model with no $\Lambda$ (right
panel). In both cases the fiducial model has $\Omega_0 = 0.3$, with
$\omegal = 0.7$ and 0 respectively. Solid lines are for neutral case
SNAP errors while dotted (dashed) lines are for best (worst) case ones.}
\end{figure}

\begin{figure}[p]
\psfig{file=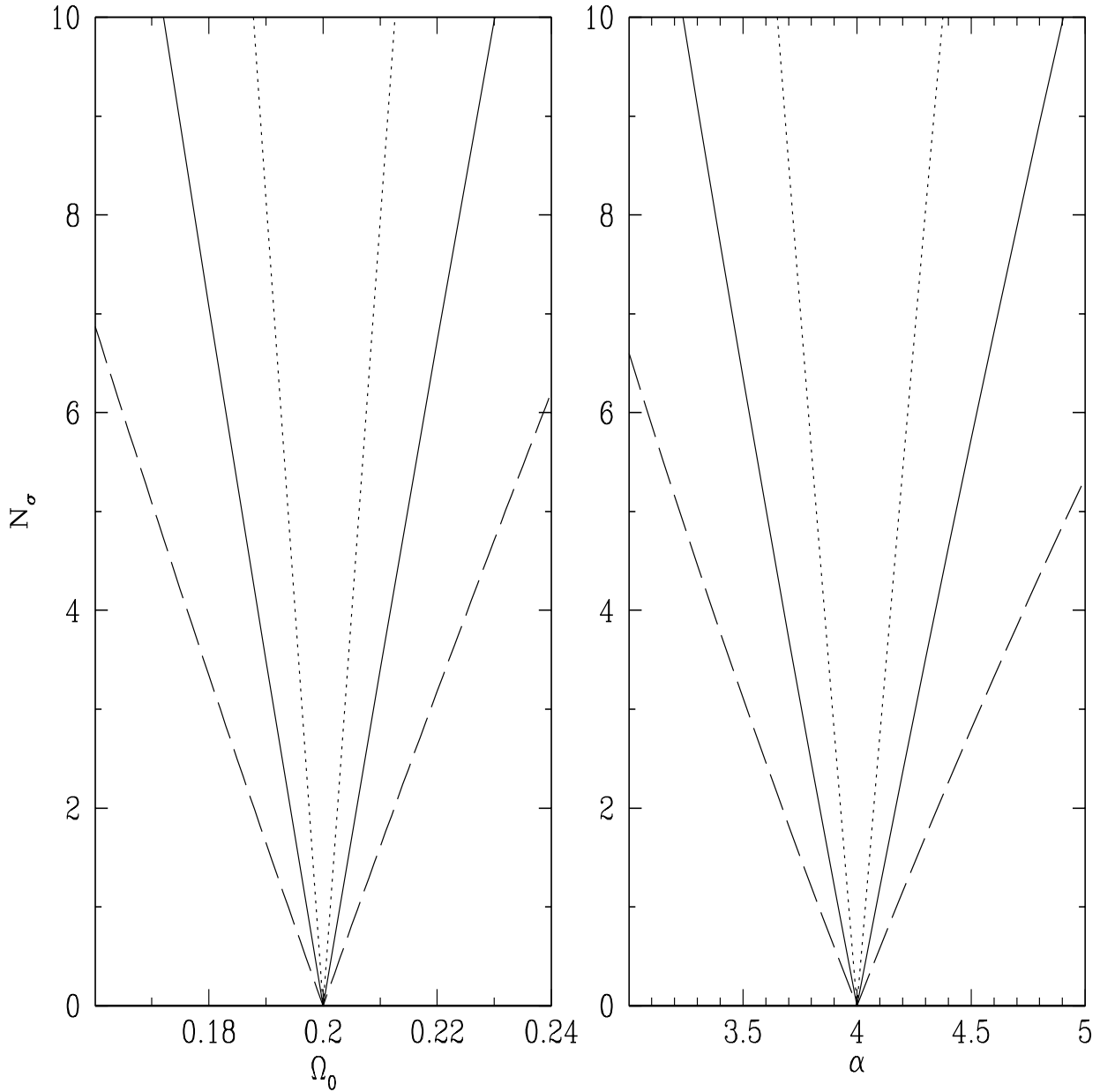,height=7.0in,width=7.0in,angle=270}
\caption{$N_\sigma (\Omega_0)$ (left panel) and
$N_\sigma(\alpha)$ (right panel) for the spatially-flat time-variable
$\Lambda$ model (Peebles \& Ratra 1988). In both cases the fiducial
model has $\Omega_0 = 0.2$ and $\alpha = 4$.  Solid lines are for
neutral case SNAP errors while dotted (dashed) lines are for best
(worst) case ones.}
\end{figure}

\end{document}